# Material Targets for Scaling All Spin Logic

Sasikanth Manipatruni, Dmitri E. Nikonov, and Ian A. Young

Components Research, Intel Corp.,

Hillsboro, OR 97124

All-spin logic devices are promising candidates to augment and complement beyond-CMOS integrated circuit computing due to non-volatility, ultra-low operating voltages, higher logical efficiency, and high density integration. However, the path to reach lower energy-delay product performance compared to CMOS transistors currently is not clear. We show that scaling and engineering the nanoscale magnetic materials and interfaces is the key to realizing spin logic devices that can surpass energy-delay performance of CMOS transistors. With validated stochastic nano-magnetic and vector spin transport numerical models, we derive the target material and interface properties for the nanomagnets and channels. We identified promising new directions for material engineering/discovery focusing on systematic scaling of magnetic anisotropy ($H_k$) with saturation magnetization ($M_s$), use of perpendicular magnetic anisotropy, and interface spin mixing conductance of ferromagnet/spin channel interface ($G_{mix}$). We provide systematic targets for scaling spin logic energy-delay product toward a 2 aJ.ns energy-delay product, comprehending the stochastic noise for nanomagnets.

AUTHOR EMAIL ADDRESS: sasikanth.manipatruni@intel.com

**CORRESPONDING AUTHOR FOOTNOTE:**

[*]To whom correspondence should be addressed. E-mail: sasikanth.manipatruni@intel.com

The control and manipulation of nanoscale spin transport and magnetism is one of the most promising approaches for beyond CMOS-logic, memory and analog applications [1, 2, and 3]. Spintronic logic devices [4-10] are promising candidates for scaling the energy efficiency of computing devices due to the promise of 1) non-volatility [2]; 2) higher logical efficiency [11]; and 3) high density integration [12, 13]. One leading device, the *All Spin Logic* (ASL) device [5, 6, 9, 11], processes information using only spin for the signal state variable, thus avoiding repeated state variable conversion and making it an excellent candidate for spin logicfor computation. Impressive experimental [14-17] breakthroughs have been achieved in the recent years accelerating the development of spintronics as a viable logic and memory technology. However, existing nanomagnet material properties and spin transfer channel properties fall short of the energy and delay targets [3] dictated by modern advanced CMOS devices [3]. Hence, it is of great interest to identify and pursue material and interface target parameters with which ASL devices can be scaled to achieve high efficiency.

In this letter, we show that scaling and engineering the nanoscale magnetic materials and interfaces is the key to realizing spin logic devices that can replace/augment CMOS devices. Using validated numerical models we derive the target material and interface properties for the nanomagnets and channels. We identify promising new vectors for material exploration focusing on a) Systematic scaling of magnetic anisotropy ($H_k$) with saturation magnetization ($M_s$); b) Use of perpendicular magnetic anisotropy; and c) Interface mixing conductance of ferromagnet/spin channel interface ($G_{mix}$).

We first describe a representative nanoscale spin logic device operating via interaction of spin currents with nanomagnets. An inverting/non-inverting spin logic device with directionality of signal (information) flow is shown in Figure 1A. The spin logic device is comprised of two nanomagnets sharing a spin conduction channel. Each of the nanomagnets is shared between two spin conduction channels for propagating and regenerating the signal information. More complex circuits, such as majority gates, can be created by having more input magnets sharing the conduction channel. The all spin logic device is also amenable to 3D integration which provides an additional means for logic density scaling (Figure 1B). The example materials for forming the nanomagnets can be CoFeB, Pd, NiFe or L10 metals (FePt, FePd). Example channel materials are Ag [14], Cu [17] and Graphene [15] which exhibit excellent room temperature spin flip lengths.

The ASL device operates by the asymmetric injection of spin currents into the spin conduction channel from the nanomagnets, which are sharing the spin conduction channel. The asymmetry may be created either from different (asymmetric) overlap of the magnets with the input and output channels, or different spin injection efficiency. In this paper, we consider asymmetric overlap (Figure 1A). When the ASL device is operated with a negative (rather than positive) supply voltage, the input and output nanomagnets inject electrons into the spin conduction channel where the spin direction is decided by the orientation of the nanoscale ferromagnets (FM) (example FM with spin pointing to the +X direction injects electrons oriented in the +X direction). The dominating magnet therefore sets up a larger spin polarization in its orientation in the channel. A spin current flows from the higher spin potential [6] to the weaker magnet (FM2) creating a spin torque to orient the weaker magnet to FM1. For positive supply voltage, a converse process happens creating the FM2 to align antiparallel to the FM1.

We model the spin logic device using an equivalent vector spin circuit for the spin logic device based on vector spin circuit theory [6, 18, and 19]. The functional form of the spin matrices are arrived at using a Landauer-Büttiker formalism with spin transport [19] applied to metallic circuits. The spin equivalent circuit modeling and the conduction elements for the ferromagnet (FM) to normal magnet (NM) interface are described in Ref [6]. Figure 2 shows the result of spin circuit simulations of the spin logic device with coupled stochastic nanomagnet dynamics and spin transport through the metallic channel. The operating dynamics of the device for nominal material conditions, with a positive applied voltage is shown in Figure 2. Figure 2A, shows the response of the device with a 10 mV positive supply voltage. The Joule power dissipated by the supply is shown in Figure 2B. The large quiescent power due to the metallic device nature can be observed which shows the need for novel power management to take advantage of the non-volatility to reduce this leakage power. The injected spin current in the +/-X direction is shown in Figure 2C. Figure 2D shows the vector equivalent circuit for the spin logic device where the magnets and the channels are modeled with 4X4 tensor conductance elements. The vector spin conductances $G_{FM1}$ and $G_{FM2}$ describe the spin conduction through the magnets into the metallic spin channel; the spin conduction conductances $G_{SeT}$, $G_{sfT}$ describe the conduction through the spin channel. We represent the metallic spin channel between node 1 and 2 as a combination of two T-equivalent circuits. The conductance $G_{sfT}$ models the loss of spin current due to spin flip in the channel. The $G_{sfT}$ is such that no charge current flows into the ground (open circuit for charge currents) creating a virtual ground for spin currents only. The nanomagnet conductance is modeled by the spin conductance tensors $G_{FM1}(\hat{m}_1)$, $G_{FM2}(\hat{m}_2)$. The spin conductance tensors for the nanomagnets vary according to the vector position of the magnet as decided by the nanomagnet dynamics.

Thermal noise plays a major role in modulating the direction of the nanomagnets magnetic moment vector at any given time. It is therefore important to model the coupled nanomagnet dynamics with stochastic effects of nanomagnets at room temperature to obtain the correct response time of spin logic. The phenomenological equations describing the dynamics of nanomagnets are (the modified Landau-Lifshitz-Gilbert-Slonczewski (LLG) equation [20])

$$\frac{\partial \hat{m}_i}{\partial t} = -\gamma\mu_0[\hat{m}_i \times \vec{H}_{eff}(T)] + \alpha\left[\hat{m}_i \times \frac{\partial m_i}{\partial t}\right] + \frac{\hat{m}_i \times (\vec{I}_{si} \times \hat{m}_i)}{eN_s} \quad (1)$$

where $\gamma$ is the electron gyromagnetic ratio; $\mu_0$ is the free space permeability; $\vec{H}_{eff}(T)$ is the effective magnetic field due to material/geometric/surface anisotropy, with the thermal noise component [21]; $\alpha$ is the Gilbert damping of the material and $\vec{I}_\perp$ is the component of vector spin current perpendicular to the magnetization ($\hat{m}$) leaving the nanomagnet, $N_s$ is the total number of Bohr magnetons per magnet. Implicit in the LLG equation is the fact that absolute values of the magnetic moments of single domain nanomagnets remain constant. The spin currents entering the nanomagnets are derived from the vector admittance of the circuit calculated based on the present conditions of the magnets.

$$\vec{I}_{ij} = G_{ij}(\hat{m}_1, \hat{m}_2)\Delta\vec{V}_{ij} \quad (2)$$

The dynamics of the spin device are solved self-consistently with the spin transport in the equivalent circuit models. The LLG solvers pass the condition of the magnets to the spin circuit and the spin circuit solver passes the spin vector current to the LLG solver at each pass of the self-consistent loop until a solution is reached. The thermal noise due to microscopic degrees of freedom of the nanomagnets plays a critical role in the dynamics of the nanomagnets and manifests itself as fluctuations to the internal anisotropic $H_{eff}(T)$ magnetic field [21]. A description of the noise properties and numerical methods used for stochastic LLG equations is provided in [6].

We now evaluate the energy-delay product of nominal the ASL device (comprising magnets with 40kT energy barrier corresponding to 7-year retention time) with nominal existing material choices. Materials with existing proof of concept integration in CMOS are chosen. Nominal material parameters for the materials are shown in Table 1. At nominal operating voltage of 10 mV, the ASL device operates with a response time ~0.5 ns dominated by the stochastic delay of switching where the nanomagnet is acquiring the stochastic phase to start the nanomagnet dynamics. The estimated energy-delay product of the device is given by the total joule energy supplied by the supply voltage. The energy-delay product of the in-plane ASL is 3 to 6 fJ.ns (variation is attributed to the stochastic nature of the switching). This falls short of the high performance 22nm CMOS node's fan-out of 4 (FO4) inverter energy-delay product by at-least an order of magnitude. The constant energy-delay contours of the in-plane ASL with stochastic simulation is obtained by varying the supply voltage and can be seen in Figure 3A. This is compared with the FO4 inverter based on low-power CMOS at the 22nm process node with a typical channel width (20 F, F=22 nm) shown as a benchmark reference [22], see Fig 8.a. The assumed operating conditions for the CMOS circuit are 0.5 V supply voltage, 1 fJ switching energy, 1.2 nW leakage power and 50 ps response time. Note that the energy-delay of a minimum sized FO4 inverter based on high-performance CMOS with a minimum channel width of F has energy on the order of 9.2 aJ and FO4 delay 3.8 ps [3].

Now we will describe the energy-delay reduction scaling of ASL using perpendicular magnetic anisotropy (PMA) material. The perpendicular anisotropy may be achieved by means of surface anisotropy (CoFeB with material thickness less than 1.2 nm) [23], L10 materials (FePt, FePd) [24], or by magnetic superlattices (Co/Pd) [25]. The perpendicular anisotropy significantly reduces the critical current for switching magnetic materials. We numerically simulate the effect of PMA on the ASL device and show significant reduction in its energy-delay product. We estimate it for a CoFeB PMA ferromagnet to be 170 aJ.ns using existing surface PMA materials,

Figure 3. The output nanomagnet dynamics are shown in Figure 3B and 3C comparing the nanomagnet dynamics of in-plane and PMA ASL.

We also show that spin diffusion length and Gilbert damping of the magnets have limited effects on improving the energy-delay of the spin logic devices. Above a certain spin diffusion length (when the spin diffusion length is longer than the channel length) the beneficial effect of better spin channels for logic device operation decreases, Figure 4. We describe the effect of the spin diffusion length of the channel material on the energy-delay performance of the ASL device in Figure 4A where the channel length is 100 nm. The effect of varying the Gilbert damping of the magnet is shown in Figure 4B. The Gilbert damping is related to the noise of the magnetization caused by stochastic magnetic fields. Thus we conclude that any energy-delay reduction scaling expected from spin diffusion length improvement and Gilbert damping improvement is minor compared to improving the anisotropy.

We now describe a second energy-delay scaling technique for spin logic devices by material and nanomagnet engineering where the energy-delay can be scaled by a linear trade-off of material magnetization saturation ($M_s$) with magnetic anisotropy field ($H_k$) of the nanomagnets. The effect of scaling up the anisotropy field while reducing the saturation magnetization of the nanomagnets by the same factor is shown in Figure 5. The nanomagnet volume and the effective thermal energy barrier remain constant. Hence, novel material discovery with a systematic control over the magnetization and $H_k$ can provide significant improvements in energy-delay product. A factor of 4 scaling of Ms and $H_k$ from the benchmark material (CoFeB) shows that the energy-delay product can be smaller than that for the 22 nm CMOS device by a factor of 3. This will provide active energy-delay product improvements equivalent to 8 generations of CMOS scaling (CMOS in 2028, [3]). An energy-delay product reduction is expected both for in-plane and PMA ASL. The optimized PMA-ASL provides energy-delay product comparable to the 22 nm CMOS device node (Figure 5B) with the added advantages of non-volatility and low supply voltage operation.

Next we will describe a third technique for energy-delay reduction scaling using the optimization of an interface property of the NM-FM interface - the spin-mixing conductance - to optimize the active energy-delay of ASL devices. The spin mixing conductance of the NM-FM interface is responsible for the responsivity of the nanomagnet to the applied spin voltage. The non-collinear spin current injected into the FM for an applied voltage can be written as:

$$\vec{I}_{S\perp} = G_{SL} \Delta \vec{V}_{\perp} \tag{3}$$

where $G_{SL}$ is the spin mixing conductance of the NM-FM interface. The spin mixing conductance is described in terms of spin reflection at the interface [19] as:

$$G_{SL} = 2\operatorname{Re} G^{\uparrow\downarrow} = \frac{2e^2}{h} \sum_{n \in NM} \left(1 - \sum_{m \in NM} r_{\uparrow}^{nm} r_{\downarrow}^{nm*}\right) = \frac{2e^2}{h} n = \frac{2e^2}{h} \frac{A k_f^2}{4\pi} \tag{4}$$

where $e^2/h$ is the conductance per spin of a ballistic channel with ideal contacts [19]; $r_{\uparrow}^{nm}, r_{\downarrow}^{nm}$ are the reflection coefficients of the up and down spin electrons at the FM-NM interface; $n$ is the number of modes in the NM, $m$ is the number of modes in the FM. The number of modes in a normal metal $NM$ can in-turn be written from the metal's Fermi wave vector $k_f$ [26]. It has been argued that $r_{\uparrow}^{nm}, r_{\downarrow}^{nm}$ are close to zero for many material systems [19], which simplifies the spin torque conductance to $G_{SL}$ as shown in (4). A NM-FM interface is shown in Figure 6A where a vector spin current (comprising of spin with a general orientation with respect to to the magnets) can be injected into the device. Figure 6B, shows a 4-component spin conductance tensor matrix for a NM-FM interface where the vector spin current is generated from the applied vector spin voltages to the NM-FM combination. Typical spin mixing conductance is set by the number of available modes in the spin channel. The effect of spin mixing conductance is shown in Figure 6 where the spin mixing conductance of the channel is varied from $0.55 \times 10^{15}$ $\Omega^{-1}m^{-2}$ to $5.5 \times 10^{15}$ $\Omega^{-1}m^{-2}$. The energy-delay product of nominal in-plane ASL can be improved by 2 orders of magnitude for an order of magnitude improvement in spin mixing conductance. Hence, it is of great interest to identify feasible mechanisms to enhance spin mixing conductance using novel channel materials or interface layers.

Finally we show that scaled magnetic materials with optimized interfaces can provide significant energy-delay product improvements for ASL devices. We compare the energy-delay product of 4 scaled ASL devices to the energy-delay of low-power 22nm CMOS in Figure 7. The energy-delay performance of the ASL devices (~6.4 fJ.ns) with the nominal existing materials and CoFeB FM and Cu channels falls short of the CMOS energy-delay product as discussed earlier. However, a perpendicular material configuration such as CoFeB with surface anisotropy can significantly reduce the energy-delay product to a few hundreds of aJ.ns. With a factor 4 improvements in material parameters, see [27, 28], by scaling the $M_s$ down to $0.25 \cdot 10^6$ A/m and increasing the $H_k$ to $1.6 \cdot 10^5$ A/m, the energy-delay product of ASL may surpass future state of the art CMOS technology. For example, we note that tetragonally distorted Huesler compounds such as $Mn_3Ga$ exhibit low magnetization ($M_s=0.11 \cdot 10^6$ A/m), reasonably large spin polarization of 58% and extremely high anisotropy fields exceeding $10^7$ A/m [28]. A variety of material choices remain to be explored for spin logic and embedded memory providing the opportunity for more improvements in energy-delay as well as device density [27, 30].

To provide even further energy-delay product improvement, we propose the plausible case of an engineered interface with an enhancement in spin mixing conductance such that the energy-delay product can be reduced to as low as 2 aJ.ns, thus providing superior energy-delay product combined with non-volatile logic operation.

Finally at the logic compute level, the energy-delay product of the ASL devices is further enhanced by taking advantage of the effect of the ASL devices' non-volatility. The effective energy per cycle for CMOS logic including the energy due to the leakage power can be written as

$$E_{eff} = \alpha E + P_{Leak} \frac{1}{f} \tag{5}$$

Where f is the clock frequency, α is the net activity of the transistors in the CPU which ranges between 0.03 (for high performance computing processors) to $10^{-6}$ (for aggressively clock gated mobile processors), $P_{Leak}$ is the leakage power of the device. For a 22nm CMOS device with channel width of 400nm, we assumed a 1.2 nW leakage power ([22], see Fig 8.a). We assumed a zero leakage power for ASL devices ignoring the overhead $CV^2f$ power needed for supply clocking/gating to off. The effective energy per bit with leakage power is plotted in Figure 7 for ASL devices with various magnetic materials. The in-plane ASL devices with nominal materials are not competitive to state-of-art CMOS, at high activity factors. However, improved energy-delay product using the introduced material scaling methods, combined with exploiting of non-volatility, can lead to better effective energy-per -bit at lower activity factors. In conclusion, we outline a scaling path for development of materials and nanomagnetic interfaces for spin based logic using lateral spin valves for higher efficiency spin logic devices. Given the strong need for high efficiency mobile and high performance computing, 3D non-volatile spin logic with scaled magnetic materials can enable radical improvements in computing throughput and energy efficiency.

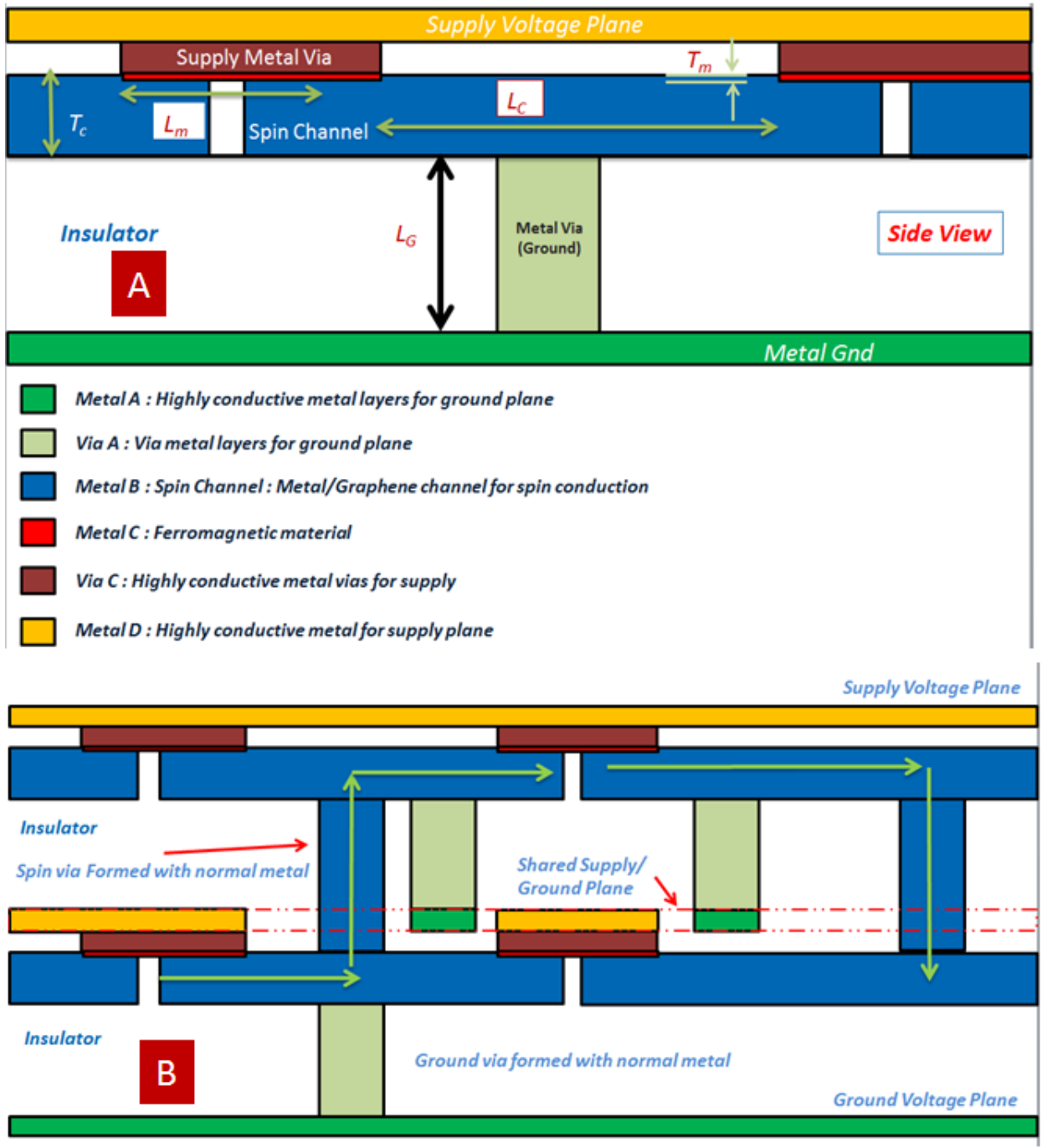

Figure 1 A) A lateral spin valve formed with two free ferromagnetic layers with directional logic operation (inverting/non-inverting gate). B) A stacking scheme for 3D all metallic All Spin Logic. Nonmagnetic vias allow for spin conduction out of plane creating a 3D stackable logic. A possible spin information flow out of plane is identified.

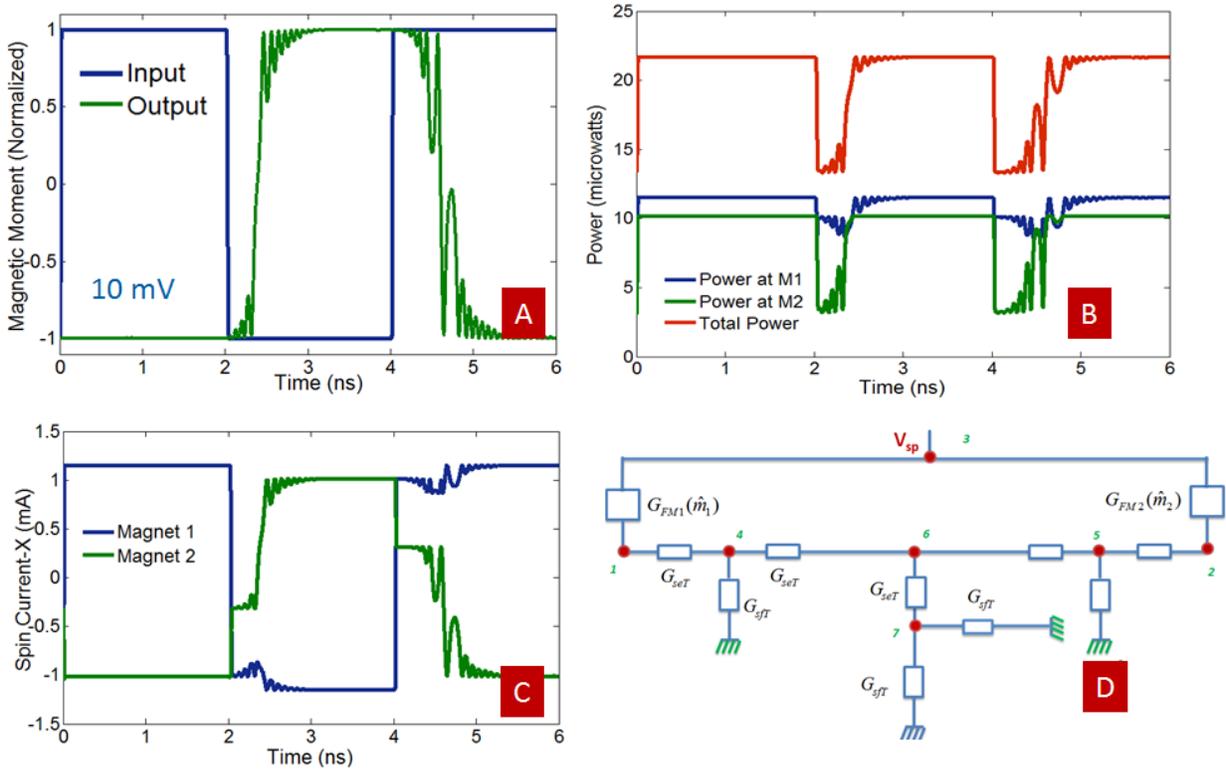

Figure 2 A) Nominal operating condition of a spin logic inverter (with a 10 mV supply voltage) B) Electrical power injected into the device at the supply terminals is shown based on a vector spin circuit model. Nominal material properties are assumed for the baseline operation of the device. C) Spin current along the X direction (direction of channel length) injected at the magnets. D) Equivalent spin circuit model for lateral spin logic device.

| Table 1. Nanomagnet and transport parameters for LLG [6] | | | |
|---|---|---|---|
| **Variable** | **Notation** | **Value/Typical Value** | **Units (SI)** |
| Free Space Permeability | $\mu_0$ | $4\pi \times 10^{-7}$ | $JA^{-2}m^{-1}$ |
| Gyromagnetic ratio | $\gamma$ | $17.6 \times 10^{10}$ | $s^{-1}T^{-1}$ |
| Saturation Magnetization of the Magnet | $M_s$ | $10^6$ | A/m |
| Damping of the Magnet | $\alpha$ | 0.007 | - |
| Effective Internal Anisotropic Field | $H_{eff}$ | $3.06 \times 10^4$ | A/m |
| Barrier of the magnet | $\Delta/kT$ | 40 | |
| Length of Magnet | $N_s$ | $10^3 - 10^6$ | - |
| Thickness of Magnet | $T_m$ | 3 | nm |
| Width of Magnet | $W_m$ | 37.8 | nm |
| Length of Magnet | $L_m$ | 75.7 | nm |
| Length of channel | $L_c$ | 100 | nm |
| Thickness of channel | $T_c$ | 200 | nm |
| Length of ground lead | $L_g$ | 200 | nm |
| Thickness of ground lead | $T_g$ | 100 | nm |
| Channel conductivity | $\rho$ | $7 \times 10^{-9}$ | $\Omega \cdot m$ |
| Sharvin conductivity | $G_{sh}$ | $0.5 \times 10^{15}$ | $\Omega^{-1} \cdot m^{-2}$ |
| Polarization | $\alpha_c$ | 0.8 | |

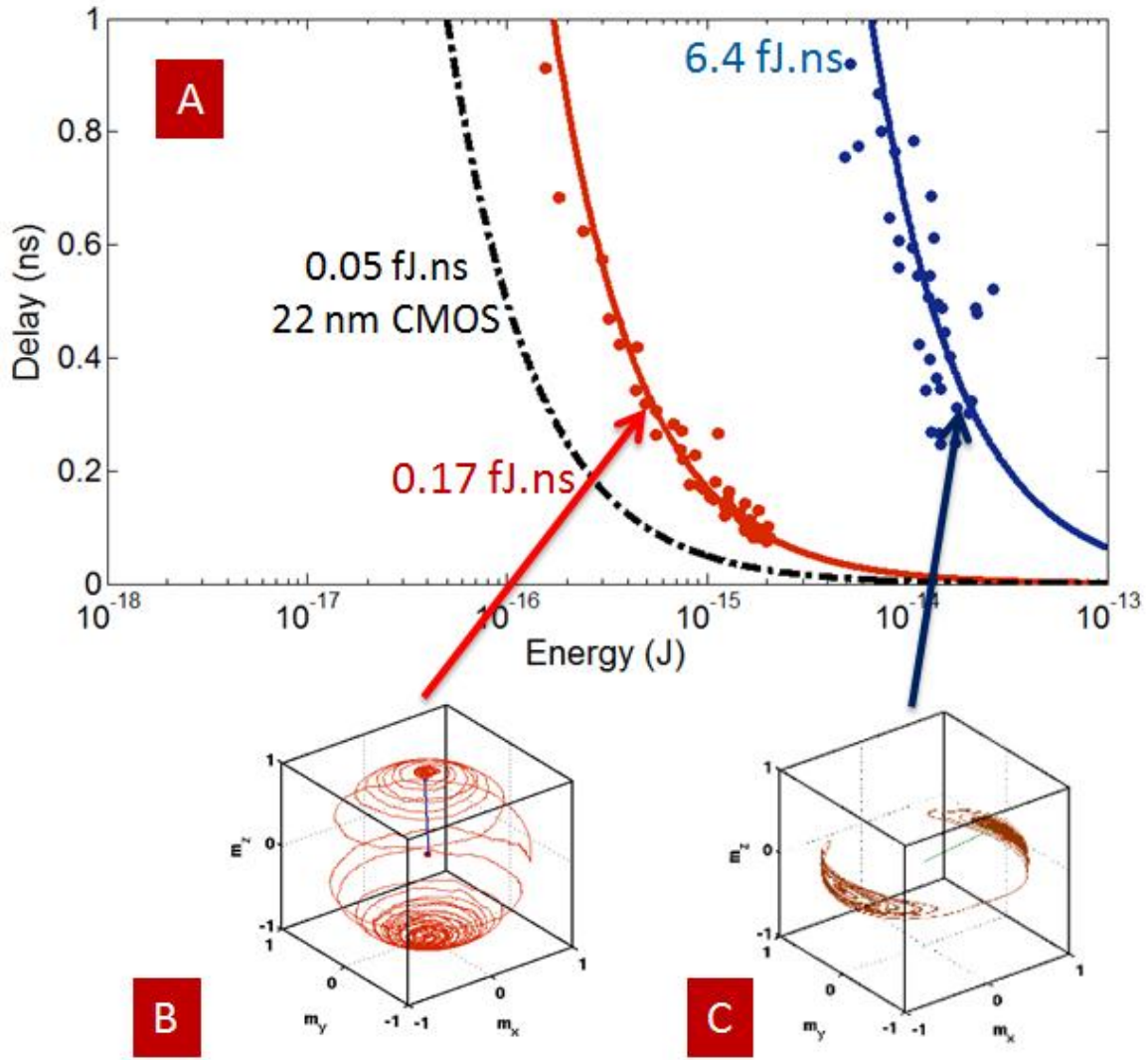

Figure 3 A) Energy vs. delay comparison of ASL devices with nominal materials using in-plane and PMA nanomagnets C) Nanomagnet dynamics with in-plane magnetization B) Perpendicular anisotropy nanomagnets.

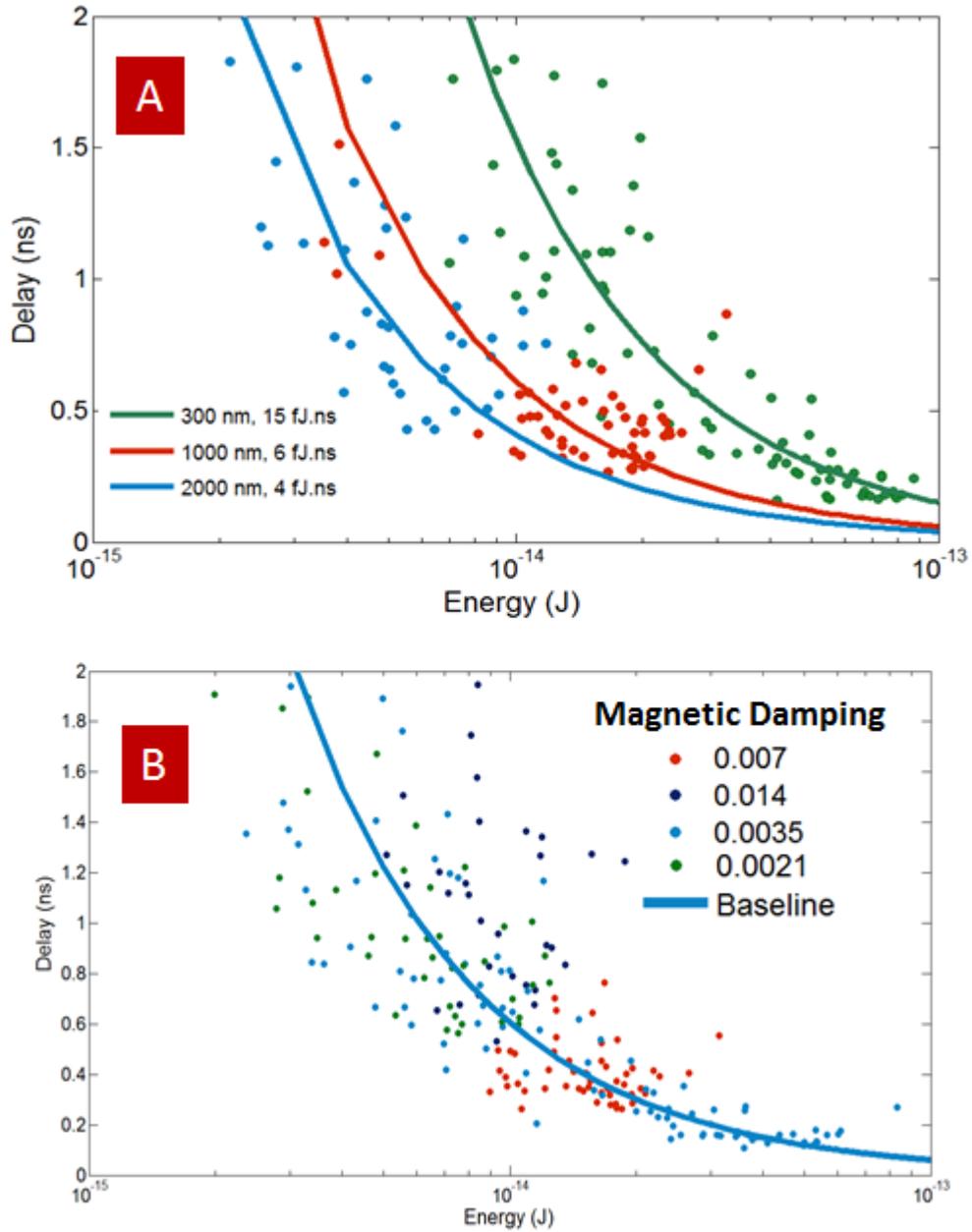

Figure 4A) Effect of spin diffusion length on the energy vs. delay of all spin logic device B) effect of Gilbert damping on the energy vs. delay of all spin logic device.

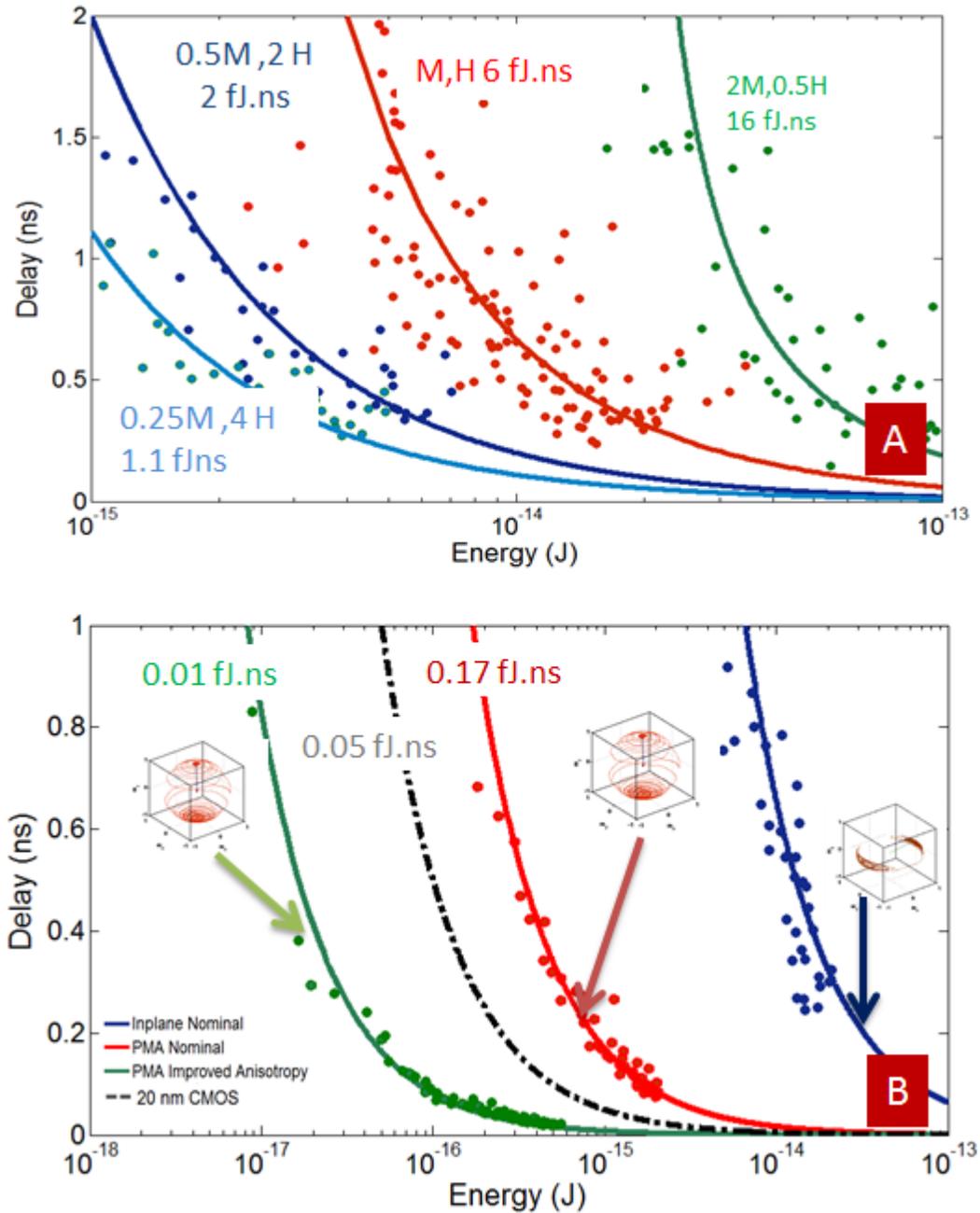

Figure 5A. Energy vs. delay comparison of in-plane ASL devices with nominal materials (red line) and scaled magnetic materials. An approximately linear order energy-delay improvement can be expected with higher $H_k$ and lower $M_s$. B. Energy vs. delay of PMA-ASL, in-plane ASL and improved PMA ASL (green) calculated via vector spin modeling of the ASL.

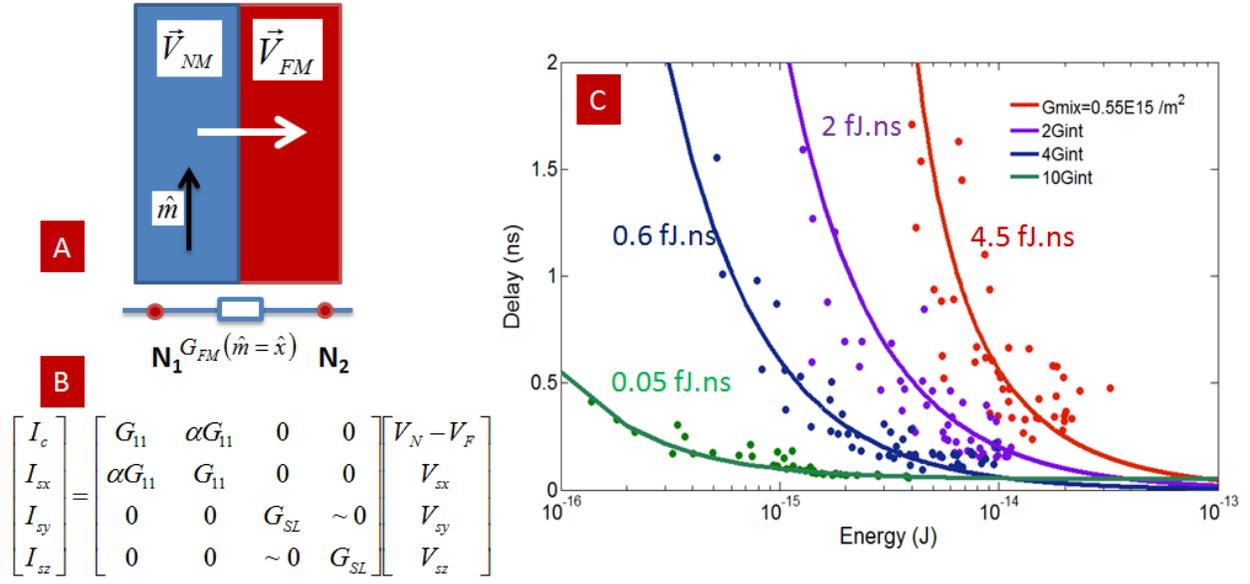

Figure 6A. C) Effect of spin mixing conductance on energy vs. delay of in-plane ASL devices with nominal materials (red line) and interfaces with enhanced spin mixing conductance.

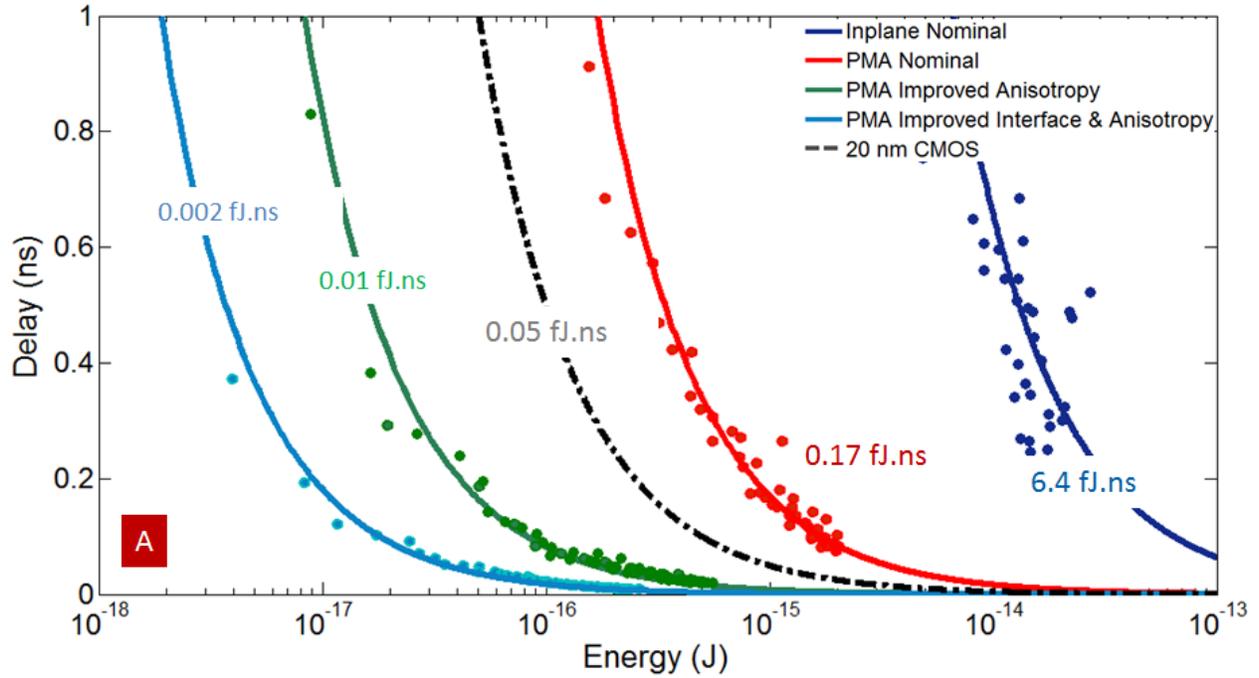

| B | | In-plane | PMA | Improved PMA | Improved PMA + Interface |
|---|---|---|---|---|---|
| **Energy-delay** | aJ.ns | 6400 | 170 | 10 | 2 |
| $M_s$ | (A/m) | $10^6$ | $10^6$ | $250 \times 10^3$ | $250 \times 10^3$ |
| $H_k$ | (A/m) | $3 \times 10^4$ | $4 \times 10^4$ | $16 \times 10^4$ | $16 \times 10^4$ |
| $G_{spin-mix}$ | ($\mu m^{-2} \Omega^{-1}$) | 550 | 550 | 550 | 2200 |

Figure 7A. Energy vs. delay of nominal in-plane ASL, PMA ASL, Scaled H-K PMA ASL and scaled PMA ASL with enhanced spin mixing conductance. B) Nanomagnet magnetic material and interface property targets for various projected devices.

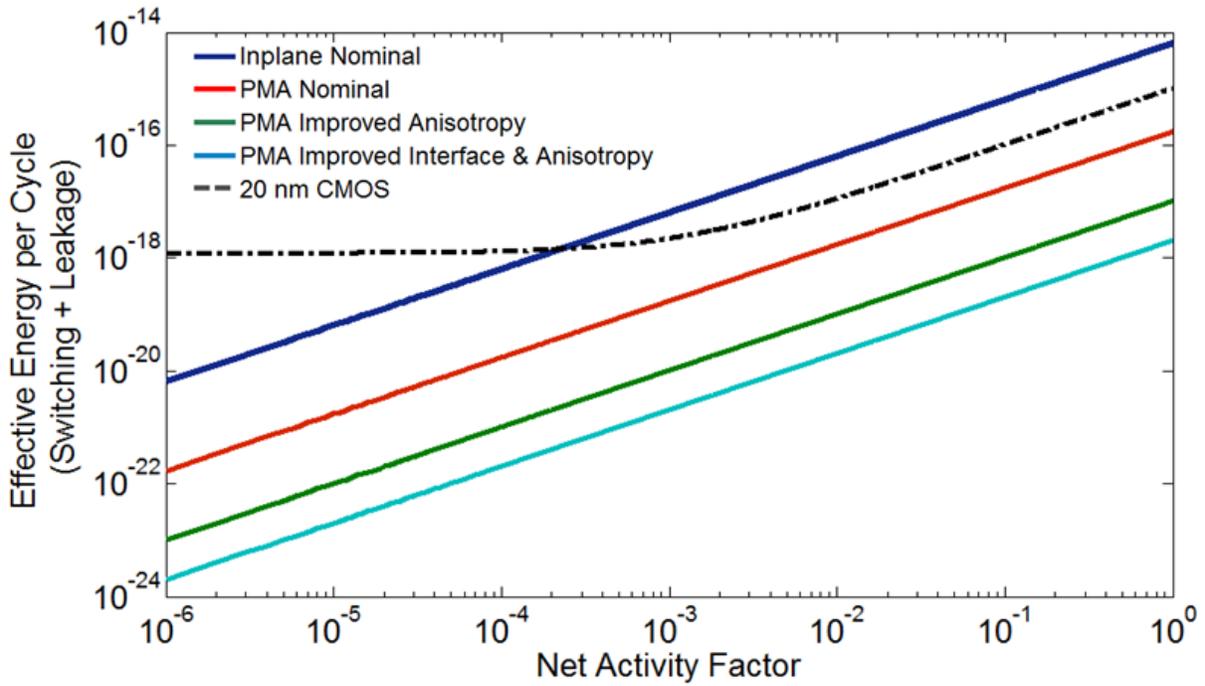

Figure 8. Effective energy per cycle including leakage power at 1 GHz clock rate. The 22 nm CMOS line assumed no active power management. The effect of non-local memory fetching & interconnects are not comprehended. Non-volatile logic can provide effectively low energy per bit compared to CMOS at low activity factor utilization.